\documentclass[11pt]{article}
\usepackage{geometry}                
\usepackage{graphicx}
\usepackage{amssymb}
\usepackage{epstopdf}
\usepackage[table,xcdraw]{xcolor}
\usepackage{booktabs}
\usepackage{lscape}
\usepackage{longtable}
\usepackage{float}
\usepackage{pdfpages}
\usepackage{authblk}
\usepackage{url}
\DeclareGraphicsExtensions{.pdf,.png,.jpg}
\title{Challenges and Considerations for Utilizing Burst Buffers in High-Performance Computing}
\author[1]{Melissa Romanus\thanks{melissa.romanus@rutgers.edu}} 
\author[2]{Robert B. Ross\thanks{rross@mcs.anl.gov}}
\author[1]{Manish Parashar\thanks{parashar@rutgers.edu}}
\affil[1]{Rutgers Discovery Informatics Institute, Rutgers University, Piscataway, NJ}
\affil[2]{Mathematics and Computer Science Division, Argonne National Laboratory, Argonne, IL}
\date{\today}                                           

\begin{document}
\maketitle

\begin{abstract}
As high-performance computing (HPC) moves into the exascale era, computer scientists and engineers must find innovative ways of transferring and processing unprecedented amounts of data. As the scale and complexity of the applications running on these machines increases, the cost of their interactions and data exchanges (in terms of latency, energy, runtime, etc.) can increase exponentially. In order to address I/O coordination and communication issues, computing vendors are developing an intermediate layer between compute nodes and the parallel file system composed of different types of memory (NVRAM, DRAM, SSD). These large scale memory appliances are being called `burst buffers.' In this paper, we envision advanced memory at various levels of HPC hardware and derive potential use cases for how to take advantage of it. We then present the challenges and issues that arise when utilizing burst buffers in next-generation supercomputers and map the challenges to the use cases. Lastly, we discuss the emerging state-of-the-art burst buffer solutions that are expected to become available by the end of the year in new HPC systems and which use cases these implementations may satisfy.
\end{abstract}

\section{Introduction}
Advanced levels of memory hierarchy, also called ``burst buffers'', have the capacity to enhance and accelerate scientific applications on high-performance computing infrastructures. However, in order to utilize burst buffers at different levels throughout the high-performance computing system, there are a number of challenges that must be addressed. In this document, we highlight specific areas of concern for utilizing advanced memory hierarchies and illustrate the focal points of burst buffers as they apply to specific use cases. This document is not meant to propose solutions to these challenges, but rather to point out where they exist in the system.

\begin{figure}[p]
\centering
\includegraphics[width=\linewidth]{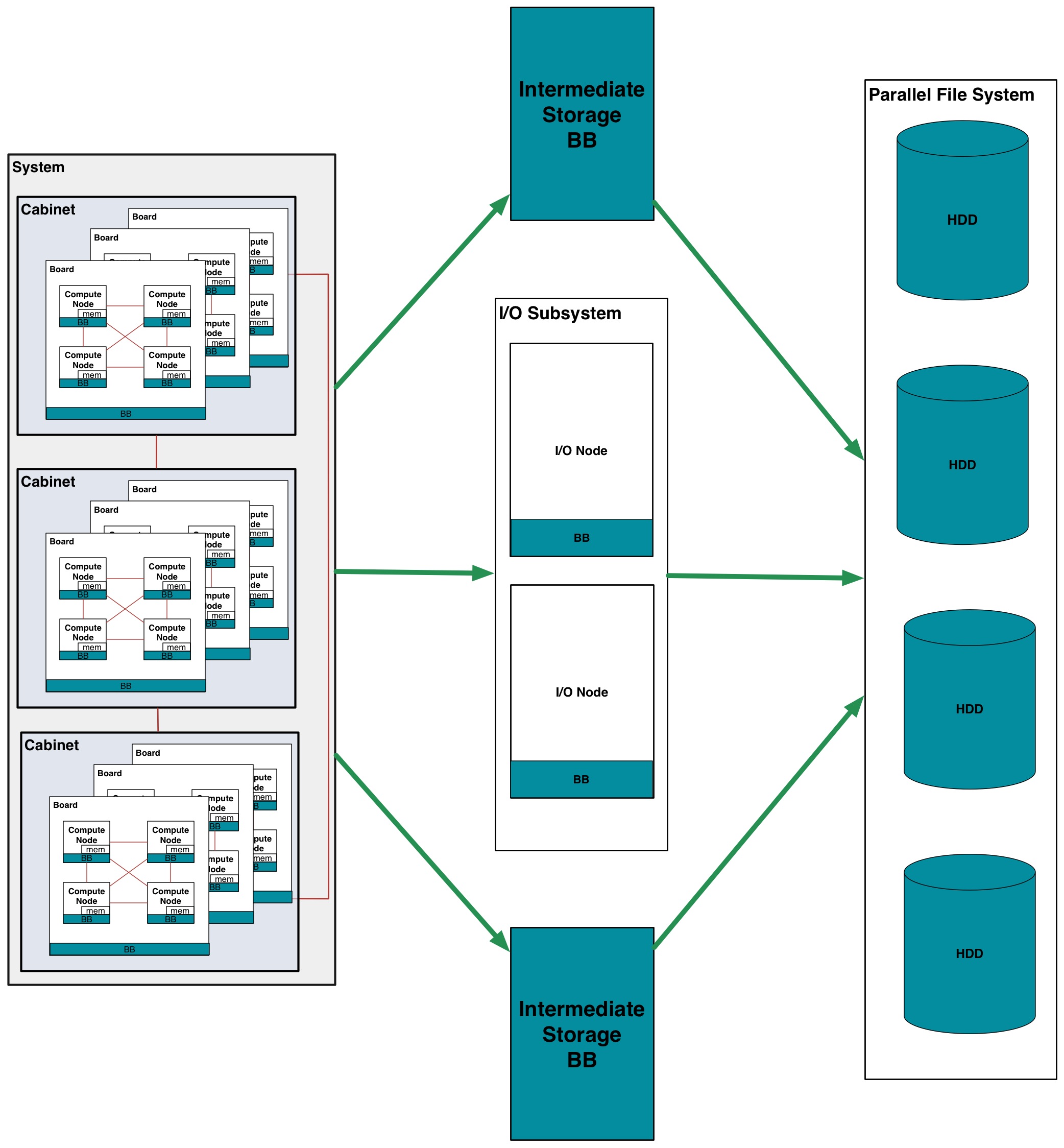}
\caption{Abstract Machine Model}
\label{fig:AMM}
\end{figure}

In order to understand the High-Performance Computing (HPC) system with burst buffer capabilities, we formed an Abstract Machine Model (AMM). We based our initial model on the existing architecture of a traditional HPC resource, and then added memory appliances at several layers in the model, e.g., \emph{Node-Local}, \emph{Board-Local}, \emph{Intermediate Storage}, and \emph{I/O Subsystem}, as seen in Figure~\ref{fig:AMM}. We use the word ``burst buffer'' in this document in order to have a general, consistent way of referring to memory at different levels of the system. It is important to note that different names may be given to these memory appliances in the future, which should only affect the naming convention expressed herein, but not the main ideas conveyed.

\begin{figure}[!hbp]
\centering
\includegraphics[scale=0.5]{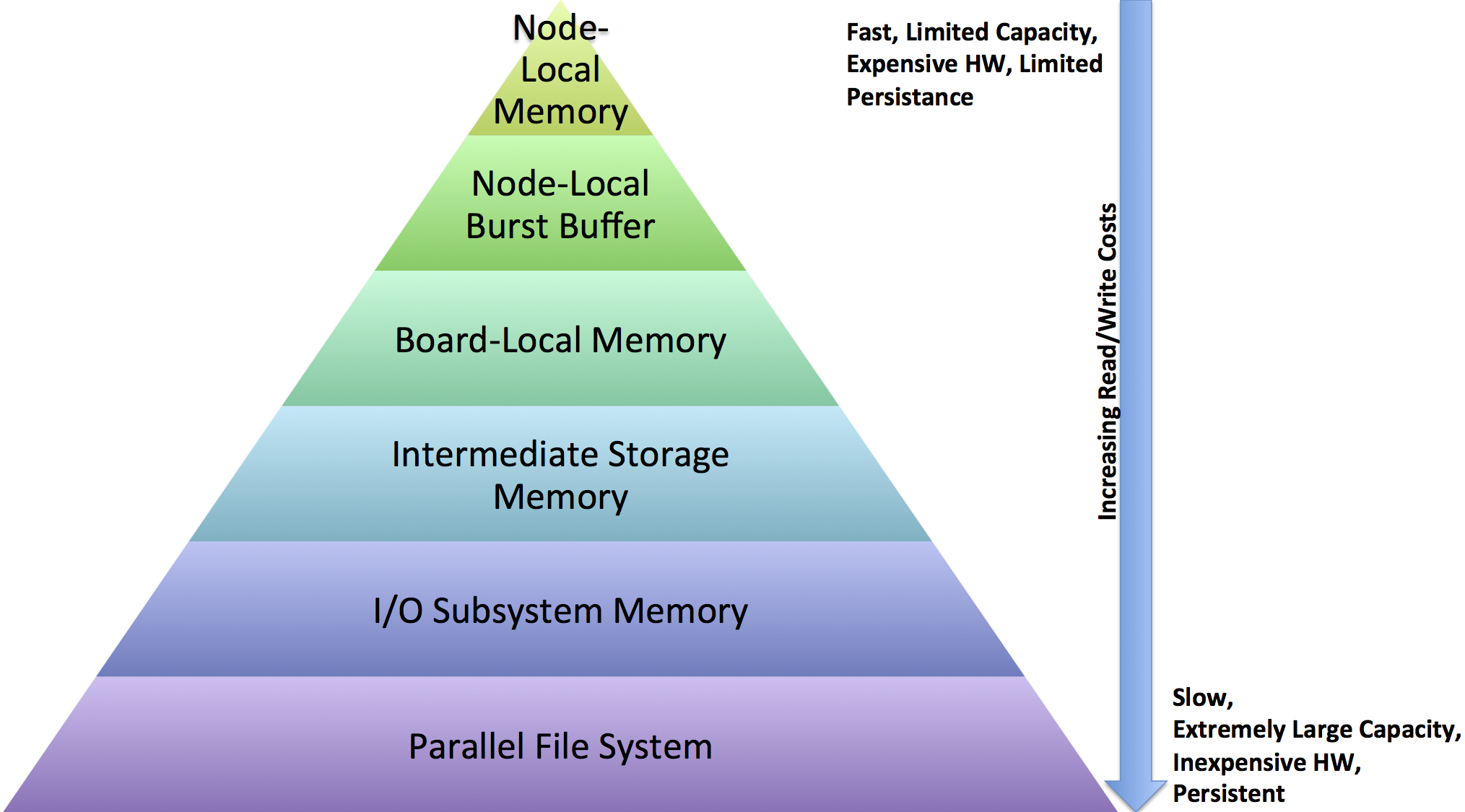}
\caption{Memory Hierarchy Model}
\label{fig:MemHier}
\end{figure}

In addition to the AMM, we present a Memory Hierarchy structure in Figure~\ref{fig:MemHier}. In this figure, memory is assumed to be available in larger quantities as you move away from the node main memory (at the top), but the cost of reading/writing to these farther away structures is much higher than reading and writing directly from and to the node memory. As part of our investigation, we have identified 3 primary areas of the HPC system that the utilization of multi-tiered burst buffers will have an impact on: (i) Applications, (ii) Programming System / Runtime Environment, and (iii) the underlying Operating System. It is important that an application or operating systems programmer is aware of the cost and overheads associated with utilizing additional memory appliances.

The rest of the document is organized as follows: Section 2 provides the motivating use cases for advanced memory hierarchies. Section 3 illustrates specific challenges and considerations that arise in offering a multi-tired burst buffer approach to support multiple applications and multiple users on an HPC system. Finally, in Section 4, we create a table that applies the general challenges identified in Section 3 to the specific use cases identified in Section 2. The specific application of the general challenges to use-case-specific scenarios provides an opportunity to identify further areas of research that must be explored in order to realize advanced memory hierarchies into future architectures.

\newpage

\section{Burst Buffer Use Cases}
\label{bbuc}
The following explains different use cases which can be enhanced by the use of burst buffers. The list of use cases is loosely based on slides from Lawrence Livermore National Laboratory~\cite{neely} and Sandia National Laboratories~\cite{leeslides}. This section aims only to discuss the data flow for each scenario, without mentioning specific locations where the data is stored. In further sections, we will discuss how and where burst buffers can be utilized, as well as what considerations apply specifically to the given use case.

\begin{enumerate}
  \item \textbf{Checkpoint-Restart}: Applications periodically write `checkpoints' of their data, i.e., snapshots of data structure or program progress. If a failure occurs (either node or data error), the applications are able to restart using a known `good' checkpoint. 
  \item \textbf{In-Situ/In-Transit Analytics/Visualization}: Data coming from simulations can be processed while it remains in the system. Also useful for visualization in the case where a node shares both a CPU and a GPU. Critical results or animations may be drained to parallel file system if they will be required at the end of the workflow.
  \item \textbf{Accelerated Reads (aka \textit{Prefetching})}: Stage data to `closest' memory in advance of when it will be read.
  \item \textbf{Out-of-Core}: The main memory on a node is not enough for the application that is currently running. As a result, advanced levels of memory hierarchy may be utilized, however, with the understanding that the `main memory' is still the fastest and `best' option.
  \item \textbf{Staged Writes (for Post-Processing)}: Files that already exist in the parallel file system (PFS) need to be processed on a compute node or a GPU at a later time than the time they were processed. 
  \item \textbf{Application/Workflow Coupling/Exchange}: Scientific workflows with coupled application components must share, exchange, modify, and communicate data to one another during runtime. 
  \end{enumerate}

\section{Burst Buffer Challenges and System-Level Considerations}

In this section, we introduce generalized `burst buffer'-specific issues that arise when considering advanced memory capable HPC systems.

\begin{enumerate}
  \item \textbf{Resource Allocation (RA)}
  	\begin{enumerate}
	\item \textit{Defining Memory Resources}: Does user on compute node automatically have burst buffer access? Does user need to define what levels of memory hierarchy s/he wants to use and how much memory s/he wants at each level? Who defines what memory resources are given to user/workflow? 
	\item \textit{Static vs. Dynamic Memory Allocation}: Is the memory allocation request fixed at the start of the job? Or is there a need to scale up or down (in terms of memory) to meet system demands and requirements?
	\item \textit{Job-Agnostic Memory Allocation}: Can memory be allocated if not associated to a particular job (i.e., prefetching or post-processing)?
	\item \textit{Charging for Varying Memory Allocations}: What is the cost for using different memory? Is it pay per allocation, pay per GB/TB, pay per use? Do different levels of storage cost different amounts of money? How to charge if more/less memory is requested during runtime? Is there a higher cost during higher demand times?
	\item \textit{Allocation Lifetime}: Does memory allocation at all levels persist for entire workflow? After application shuts down and releases CPU cores, if there is still data in memory that must be moved, should the burst buffer and node still be marked as used? How does that work with the pricing model?
	\item \textit{Burst Buffer Allocation}: How is a burst buffer allocated? In what terms is allocation performed -- i.e., is it space on a filesystem, address space, other? If multiple burst buffers are allocated to an application/workflow, what software would be responsible for the aggregate view of a distributed Burst Buffer allocation?
	\item \textit{Shared Allocation of Burst Buffers}:  If you are sharing burst buffers between users/applications, how do you define an allocation policy to decide who uses it when? Can we aggregate multiple burst buffers (of same or different type) and expose them to a multiple applications from multiple users? 
	\item \textit{Application-View of Memory}: How are different burst buffers exposed to the user/application? 
	\end{enumerate}
 \item \textbf{Access Control (AC)}
 	\begin{enumerate}
	\item \textit{Permissions}: How do we ensure only ``trusted'' applications access global data? Should user and group-based permissions (akin to file-based systems) be enforced in order to protect data? Can a workflow have a specific ID allowing it access to levels of memory with that ID? How do we partition the memory between different users and processes?
	\item \textit{Data Sharing}: Can we aggregate specific portions of memory across all levels? Can we give multiple applications access to this data? 
		\begin{enumerate}
		\item \textit{Global vs. Local Data Structures}: Can application have local data structures that cannot be accessed outside of specific burst buffer? Where in memory should global structures be stored to allow fast access by all nodes involved in workflow? 
		\end{enumerate}
	\item \textit{Fair-Use of Memory Appliances}: How is memory usage controlled, e.g., to prevent a user from requesting all of the Intermediate Storage, leaving other applications with nothing? Is there a maximum limit (in terms of time or space) an application can request at each level of memory hierarchy?
	\item \textit{Application/Workflow Connectivity}: How does a workflow or application obtain access to a given memory allocation? Is it exposed as a block of a filesystem? Does it follow shared-memory models?
	\item \textit{Data Privacy}: How do we ensure that data that was once stored on burst buffer resources is no longer available once an application/workflow has deallocated the resource (e.g., encrypted storage of data)?
	\end{enumerate}
\item \textbf{Priority (PR)}
 	\begin{enumerate}
	\item \textit{Allocation Priority}: Can we define policies for the shared access between burst buffers? For example, do we provide the ability for a node or an application to access another node's on-node burst buffer? Does application running on compute node have full priority to on-node burst buffer? Do use cases have different priorities to specific levels of memory based on their importance to overall application/workflow? Can an already allocated memory space be overtaken by a higher priority user/application? If so, how does system adjust? Can two different users share an allocation on a single compute node? For example, if one application is shutting down and using half of it, can pre-staging for the next application occur in the other half of the burst buffer? How do we keep track of what usage mode (use case) we are using in order to enforce priority? What component enforces the priority?
	\item \textit{Higher Priority as Billable Feature}: Should we allow machine users willing to pay more a greater priority over certain memory appliances, possibly closer to their computation? 
	\item \textit{Eviction}: What happens to data when a higher priority use case or allocation takes over? Can system get rid of unused data, or should all data be evicted to another memory location (possibly further away)?
	\item \textit{Multiple Users Share Burst Buffer}: Does application that is running always have highest priority to Node-Local burst buffer? If system allocates a Prefetch allocation to a Node-Local burst buffer where a different computation is running, what happens if the running application needs the additional space back? Does system enact eviction to next highest memory layer? (See also: Access Control)
	\end{enumerate}
\item \textbf{Persistence (PST)}
 	\begin{enumerate}
	\item \textit{Data Lifetime}: How long does data last after an application or workflow shuts down? Is there a time limit before data can be deleted? How does user indicate what data it would like to send to permanent file if it first goes in to a different level of memory? 
	\item \textit{Flush after Application/Workflow Ends}: What happens to data that is left in a burst buffer after application or workflow shut down? Should it just be immediately deleted since the workflow is over and one can assume if it was not consumed, it is not relevant? Or does it need to be flushed to a permanent storage solution, such as parallel file system? Who is responsible for transferring the data to the parallel file system? How is it stored -- does all data in memory get written to a specific directory in the user's home directory?
	\item \textit{Garbage Collection}: How often should system search burst buffers for dirty bits/memory and clear them? Does system need to keep track of all levels (and locations) of memory that are touched by each application or workflow? After workflow shut down (and any flush to permanent file system), does the system move systematically through each level of memory to clean up data associated with the application or workflow? Is this triggered at application shutdown or periodically or when storage is close to filling up?
	\end{enumerate}
\item \textbf{Consistency (CON)}
 	\begin{enumerate}
	\item \textit{Multiple Read/Write Requests to Burst Buffer}: Does each burst buffer need multi-threaded controller code in order to handle large magnitudes of read/write requests? 
	\item \textit{Burst Buffer Guarantees}: What consistency policies will user be guaranteed using burst buffers? Do these policies differ at different burst buffer levels? How will we ensure burst buffer constraints are not violated? How do we ensure the correctness and validity of transactions between applications, burst buffers, and file system?
	\end{enumerate}
\item \textbf{Coordination (C)}
 	\begin{enumerate}
	\item \textit{Data Movement}: How will data be transferred throughout the different layers of memory? Can we quantify the overheads for transferring between levels? What system will be responsible for transporting data using RDMA?
	\item \textit{Access Coordination}: What facilities will be provided by burst buffers for coordination of access (e.g., shared mem/token exchange)? What facilities will be provided by higher software layers (e.g., locking)? 
	\item \textit{Coordination of Burst Buffer}: How to coordinate/compose multiple burst buffers belonging to same application? How to expose the composition of burst buffers to user? How to decide which physical burst buffer to store data in when multiple burst buffers exist? 
	\end{enumerate}
\end{enumerate}

\section{Table}

Please see the following pages for a table applying the Section 3 considerations to the Section 2 use cases.



| Use Cases | Application | Programming System/Runtime | Operating System |
|---|---|---|---|
| **Checkpoint-Restart** | 1) RA: User can define initial static memory allocation size, number of checkpoints (or data size) to keep on one node<br>2) RA: Application-View of Memory - Write checkpoints to burst buffer as if it is the same as main memory | 1) RA: Allocation Lifetime - if application shuts down and checkpoint data in burst buffer, should it be flushed to PFS or deleted?<br>2) C: If checkpoint data exceeds BB capacity, runtime will have to move it to different memory level | 1) PR: How many checkpoints to store in node-local burst buffer before eviction to higher level of memory<br>2) RA: OS may have to dynamically adjust initial allocation if Node-Local memory is exhausted or higher PR use case needs to use Node-Local memory<br>3) AC: Checkpoint variables are local, does not need to share between processes/applications<br>4) PST: Garbage collection - clean up after application completes or after every n-th time step<br>5) AC: Permissions - Must ensure restricted access if two users are utilizing the same burst buffer<br>6) AC: Eviction of checkpoints when Node-Local BB is getting full or reach user maximum<br>7) PST: Does checkpoint data in BB after application shut down need to be flushed to PFS? How to control flushing mechanism - OS set flag on compute node?<br>8) CON: Multiple writes but single application after every time step makes consistency not big concern for C/R case |

| Use Cases | Application | Programming System/Runtime | Operating System |
|---|---|---|---|
| **In-situ/In-Transit Analytics/ Visualization** | 1) RA: Application-View of Memory: How to expose memory to both applications as just a persistent object store that can be read and written from?<br>2) PST: User define what data is relevant for PFS (i.e., needs to be accessed later)<br>3) RA: Defining Resources: User could define the need for Intermediate Storage or Board-Level storage for In-Transit Analysis purposes<br>4) RA: Allocation Lifetime: User may define lifetime to span workflow or to span certain applications within workflow where in-transit analysis is desired | 1) RA: Dynamic memory allocation if size of in-situ data not known at start<br>2) RA: Allocation Lifetime: When viz. or analysis will follow a simulation on same node, allocation must be held after simulation shuts down<br>3) RA: Static allocation based on user request<br>4) RA: Memory Allocation: Can intermediate storage be treated as staging area where data services are running to process the data?<br>5) CON: Coordination of Burst Buffers: If multiple burst buffers are aggregated to perform in-transit analysis, how will data be split across BBs and will it be analyzed in chunks on a per BB basis? | 1) PST: After sim. shuts down, data must persist in Node-Local BB to be consumed by in-situ<br>2) AC: Data Sharing: Data generated by sim. is local and will be consumed by next application on compute node<br>3) PR: Way to ensure that in-situ data stays as 'in-place' as possible? Where to store if in-situ data would not fit on one node?<br>4) CON: Locking on read/write necessary, sim. should not shut down until all processes finished writing to Node-Local BB, ensures viz. data is complete<br>5) PST: Data Lifetime: Is data removed from BB as soon as it is consumed in this case?<br>6) C: If BB is full of in-situ sim. data, where to place new viz. data? Coordinate data movement of semi-permanent viz. output to PFS<br>7) AC: Permissions of data access belong to the workflow<br>8) PST: Data that is analyzed can be removed after the important results have been transferred to PFS<br>9) PR: If in-transit appliance is larger, priority may not be an issue unless in times of extreme machine stress. In this case, data may be processed in blocks and then aggregated into a file to be sent to PFS<br>10) PST: Garbage Collection: Remove all data associated with in-transit analysis when allocation is up |

| Use Cases | Application | Programming System/Runtime | Operating System |
|---|---|---|---|
| **Accelerated Reads** | 1) RA: User must define to OS/Runtime that data from PFS is needed for application - so usage mode is known | 1) RA: Job-Agnostic Memory Allocation: Start prefetching data before job is running<br>2) RA/PR: Attempt to use free compute node BB, if not, running application may have priority on Node-Local BB. Possible to queue data in Board-Local BB and then transfer up to Node-Local when running application shuts down?<br>3) RA: How to ensure job is allocated where the prefetched data has been stored into BB<br>4) RA: Static Allocation because size of data that needs to be staged is known a priori.<br>5) RA: Allocation Lifetime: Can BB allocation be released after prefetched data is read?<br>6) RA: Co-Allocation of BB: If pre-fetch data is needed for multiple compute nodes, can BBs be exposed as an aggregate global memory space or can data be stored in higher memory BB that all compute nodes can easily access (minimize write transfers as well)? How to ensure BBs are physically close to where they will be used | 1) RA: How to assess cost of using BB and resources before using on-node CPU time<br>2) C: How to coordinate movement of the PFS data to assorted BB?<br>3) RA: Memory allocation: Movement of file to different type of memory appliance<br>4) AC: Permissions/Data Sharing: Will need to ensure only trusted applications or nodes can read data<br>5) PR: Prefetching usage mode must be known a priori so OS can make decisions related to use case<br>6) PR: Multiple Users Share BB: If prefetch data is stored into Node-Local BB where separate application is running, if separate application needs to store data, how to handle?<br>7) PST: Can prefetched data be removed from BB as soon as it is consumed? |

| Use Cases | Application | Programming System/Runtime | Operating System |
|---|---|---|---|
| **Out-of-Core** | 1) RA/AC: Application-View/Connectivity: When additional levels of memory are needed, how is this exposed to the application? Will application have knowledge of cost of accessing further away appliance, or will it look like one space? Will closer appliances be used as cache and further away as main memory? | 1) RA: Dynamic Allocation: Application does not know when it will need more memory or how much more memory it will need. 2) RA: Allocation Lifetime: Memory allocation will need to scale up or down as needed. Can allocation lifetime be based on demand? 3) RA: Pricing: Should Dynamic Bursting cost more, because it was an unexpected use of power/energy of the system? | 1) AC: Fair-Use of Memory Appliances: How to ensure that a bug in an application does not activate out of core and then chew through all available system memory unnecessarily? 2) AC: Permissions: How to ensure that this extension of compute node memory is only accessible to a given application on a given compute node? 3) PR: If application will crash if it is out of core, what should the priority be of this use case in relation to the others? 4) C: How to handle multiple levels of memory being used as one contiguous block? 5) PST: Garbage Collection: System needs to keep track of all memory locations used for out-of-core and flush them if allocation is decreased 6) CON: Locking on read/write is necessary, to ensure that all data arrives at destination. However, out-of-core will be reserved to single compute node so all processes making read/write requests will belong to same application |

| Use Cases | Application | Programming System/Runtime | Operating System |
|---|---|---|---|
| **Staged Writes (for Post-Processing)** | 1) RA: User define resources since size of files is known a priori<br>2) AC: Post-processing is sometimes performed by different members of a team. User responsibility to ensure filesystem permissions permit the initial read of such data | 1) RA: Static Memory Allocation of Initial Movement of Data into Burst Buffer, however, as data is processed, memory allocation requirements may become more dynamic to store processed data somewhere<br>2) RA: Job-Agnostic BB Allocation: Want to start moving data from PFS as close to compute nodes where it will be processed as possible<br>3) Please see Accelerated Reads for other relevant information | 1) Please see all of Accelerated Reads, OS. This is a similar case, except that the data will be analyzed at destination instead of just consumed. |

| Use Cases | Application | Programming System/Runtime | Operating System |
|---|---|---|---|
| **Application/ Workflow Coupling/Exchange** | 1) C: Coordination of BB: User/application must have a way of communicating with other components/applications, either through levels of memory, data staging, etc.<br>2) RA: Defining Memory Resources: User may provide application coupling semantics which could be used to drive BB coupling semantics | 1) RA: Memory Allocation: How to co-allocate burst buffers for use between coupled applications?<br>2) RA: Static allocation if application coupling is known a priori<br>3) RA: Allocation Lifetime: Depending on coupling scheme, allocation pattern may change over time if there will be periods with no coupling<br>4) RA: Memory Allocation: Applications can malloc memory or exchange via data staging | 1) AC: Global variables: Share data between applications as global and private data (that does not need to be exchanged) can remain local to compute node<br>2) AC: Permissions: How to ensure applications that are part of workflow can access data associated with the workflow?<br>3) PR: Allocation Priority: If some workflow components are already running and writing to BB, priority to BB should be given to the rest of the applications comprising the workflow.<br>4) CON: Locking for global and local data is very important in a coupling scenario, to preserve versioning information, avoid overwriting data, and prevent tasks from processing the same regions<br>5) PST: Data Lifetime: Data comprising a workflow must persist until all components of the workflow intending to use that data have completed. This means data cannot be erased after one component of workflow completes.<br>6) PST: Flush: If workflow ends and data is still in BB, how to determine if this data is relevant to be stored to PFS or if it should be deleted?<br>7) PST: Garbage collection: Triggered after flush of data, must clean any BBs that workflow accessed<br>8) CON: Multiple Read/Write Requests to BB / Will requests need to queue if maximum threads are reached?<br>9) C: Data Movement: Movement from different levels of memory must be coordinate based on the workflow access patterns<br>10) C: Coordination of Burst Buffer: How can multiple levels of burst buffers be exposed to workflow? How does OS decide best physical location or memory layer to store data in the presence of multiple BBs? |



\section{State of the Art}

In this section, we discuss the burst buffer systems developed by two top High-Performance
Computing vendors, Cray and Data Direct Networks (DDN). Both have similar hardware implementations
in that the extra memory appliance they have created is comprised of solid-state drives (SSD) and acts
as an intermediate layer between the compute nodes and the underlying file system. In these first-generation
implementations, burst buffers are exposed to the end user as mountable filesystems, or, alternatively,
I/O is autonomically managed by the underlying controller or operating system. As burst buffers become more widely adopted, it may be beneficial to expose more features to the end user and/or software developers in order to fully utilize their capabilities in the use cases mentioned in Section~\ref{bbuc}.

Cray DataWarp~\cite{dwarp} utilizes flash SSD I/O blades with the Aires high-speed interconnect network. 
In DataWarp, the burst buffer can be used as a global storage cache for the parallel file system. In this usage
mode, its focus is on I/O acceleration and optimization across the machine. According to Cray, DataWarp users can
allocate the type and amount of data storage they need, in addition to defining the I/O movement on a per job, per process,
per rank, or per node scale.  Using the machine queuing system, such as SLURM~\cite{SLURM}, users can make a persistent reservation of burst buffers, i.e., a separate reservation from a job reservation - the two are independent of one another in this usage mode. In the production release of DataWarp, data striping across multiple SSD nodes is possible. Storage is dynamically allocated, meaning that a user has the option to interact with their burst buffer reservation as if it were a `scratch' file system (e.g., looks the same as a mountable filesystem), or can set up a burst capability which is local to the compute nodes for faster Checkpoint-Restart (e.g., acts more like a cache). It could also be utilized by the underlying operating system to capture `bursty' application behavior in order to optimize the usage of the parallel file system (PFS) and minimize network traffic.

Of the use cases discussed in Section~\ref{bbuc}, we have identified the following areas where we envision DataWarp being utilized in scientific application workflows. The first such use case is Checkpoint-Restart. In checkpoint-restart, the Intermediate layer (burst buffer) could hold checkpoint data and only bleed the larger checkpoints to the PFS when it is most efficient and does not overwhelm the filesystem or network. Additionally, keeping the checkpoint data in the SSD rather than writing to PFS makes potential restarts faster, especially since the allocated SSD storage can persist even when a job fails. We also envision DataWarp as supporting out-of-core use cases, since it can dynamically allocate more memory in order to accommodate `bursty' applications. It can also utilize such extra memory to ensure peak performance of the PFS. Additionally, utilizing some type of `controller' program, the SSD could potentially also be used to pre-fetch data from PFS to SSD before a job starts up. From the data that has been released so far, it is unclear if the underlying operating system will support prefetching itself. However, if not, a controller program or user service could run on a compute node or in user space that initiates data transition from PFS to SSD after reserving a persistent DataWarp area but prior to a job running.

Data Direct Networks has also released its own burst buffer solution, called the Infinite Memory Engine (IME)~\cite{IME}. In contrast to DataWarp, DDN IME can be added to existing machines. It is also implemented as an intermediate array of memory between compute nodes and the filesystem. When this intermediate memory is SSD-based, DDN recommends a successful burst buffer solution would have 2 to 3 SSDs per compute node. The utilization of IME is independent of the high-speed interconnects in the machine. This means that DDN can connect to InfiniBand, Cray Aires, or other networks. In addition, IME accommodates wide ranges of compute node vendors and storage vendors, making it as modular as possible so users can select what they need for their specific existing systems. IME utilizes its own controller to manage the coordination and communication with the SSDs. Similar to DataWarp, the memory is exposed again in a `hands-off' manner to the user; it is built upon the principle that applications need not be modified in order to achieve I/O acceleration using IME and can be mounted with a filesystem view to the application. Additionally, IME automatically detects when the I/O network is flooded and captures some of the traffic in order to optimize the utilization of the high-speed interconnects. If necessary, it communicates with the PFS in order to store persistent information to files. Lastly, IME enables post-processing and visualization/analysis in multiple scientific applications by allowing the different coupled components to manipulate common datasets in real-time. Although this solution has not been fully deployed, DDN has released information that using IME the unmodified scientific application S3D~\cite{sd} (a well-known model for turbulent flow) was run with and without IME services and experienced 3 orders of magnitude I/O acceleration and 2 orders of magnitude PFS acceleration when utilizing IME~\cite{IME}.

Similarly to DataWarp, IME would be useful in the Checkpoint-Restart use case (for the same reasons as mentioned above). Because of its ability to self-manage high network traffic, it would also be suitable for out-of-core applications and workflows. It's touted ability to facilitate fast post-processing via analysis, manipulation, and workflow processing would also be extremely valuable. If intermediate data is kept in the IME storage appliance, post-processing applications would be able to read this information more quickly and only write to PFS the end results that the domain scientist is interested in. Lastly, with these current specifications, DDN IME could support different workflow/application coupling techniques, with support for ensembles, visualization, and rapid exchange of possibly common data structures. 

\section{Conclusion}

In this paper, we have proposed several use cases for the next-generation memory appliances, often called `burst buffers,' that are emerging in hardware. In the current state of the art offerings, this burst buffer is considered only an Intermediate Storage between compute nodes and PFS. However, we have discussed how similar storage appliances, mixing various memory options (e.g., DRAM, NVRAM, SSD, etc.), would be useful at several different areas of the HPC architecture. We then illustrate possible use cases for this extended memory architecture, as well as point out issues that may arise when supporting such use cases. Lastly, we included a discussion of current state of the art solutions. It is important to note that this discussion was based upon the currently available information. Some information may still be proprietary, and other vendors may release competitive memory appliance solutions. It is also important to note that the ability of current solutions (i.e., DataWarp and IME) to support some of the use cases discussed in Section~\ref{bbuc} does not necessarily mean that this Intermediate layer can be considered a complete solution. Certainly, more complex storage schemes and data scheduling can be achieved with the memory views described in Figure~\ref{fig:MemHier} and Figure~\ref{fig:AMM}. As `burst buffers' become more widely adopted in the supercomputing community, we can build a more complete picture of the direct interaction between the hardware and our scientific workflow use cases.

\section*{Acknowledgments}
The research presented in this work is supported in part by the Director, Office of Advanced Scientific Computing Research, Office of Science, of the US Department of Energy Scientific Discovery through the DoE RSVP grant via subcontract number 4000126989 from UT Battelle. The research at Rutgers was conducted as part of the Rutgers Discovery Informatics Institute ($\rm RDI^2$).

\bibliographystyle{abbrv}
\bibliography{BurstBuffer_June3_Final}

\end{document}